\documentclass[reprint, amsmath,amssymb, aps, prl,longbibliography,superscriptaddress,email]{revtex4-2}

\usepackage{graphicx}
\usepackage{comment}
\usepackage{amsmath}
\usepackage{mathtools}
\usepackage{color}
\usepackage{hyperref}
\usepackage[T1]{fontenc}
\usepackage[utf8]{inputenc}

\usepackage{soul}

\begin{document}

\preprint{APS/123-QED}

\title{Symmetry breaking and non-ergodicity in a driven-dissipative ensemble of multi-level atoms in a cavity}

\author{Enrique Hernandez}
\affiliation{Center for Quantum Science and Physikalisches Institut, Eberhard-Karls Universität Tübingen, Auf der Morgenstelle 14, 72076 Tübingen, Germany}
\author{Elmer Suarez}
\affiliation{Center for Quantum Science and Physikalisches Institut, Eberhard-Karls Universität Tübingen, Auf der Morgenstelle 14, 72076 Tübingen, Germany}
\author{Igor Lesanovsky}
\affiliation{Institut f\"ur Theoretische Physik, Universit\"at Tübingen, Auf der Morgenstelle 14, 72076 T\"ubingen, Germany}
\affiliation{School of Physics and Astronomy and Centre for the Mathematics and Theoretical Physics of Quantum Non-Equilibrium Systems, The University of Nottingham, Nottingham, NG7 2RD, United Kingdom}
\author{Beatriz Olmos}
\affiliation{Institut f\"ur Theoretische Physik, Universit\"at Tübingen, Auf der Morgenstelle 14, 72076 T\"ubingen, Germany}
\author{Philippe W. Courteille}
\affiliation{ Instituto de Física de São Carlos, Centro de pesquisa em óptica é fotônica, Universidade de São Paulo, Brazil }
\author{Sebastian Slama}
\affiliation{Center for Quantum Science and Physikalisches Institut, Eberhard-Karls Universität Tübingen, Auf der Morgenstelle 14, 72076 Tübingen, Germany }
\email{sebastian.slama@uni-tuebingen.de}

\date{\today}

\newcommand{\comBea}{\textcolor{blue}}

\begin{abstract} 
Dissipative light-matter systems can display emergent collective behavior. Here, we report a $\mathbb{Z}_2$-symmetry-breaking phase transition in a system of multi-level $^{87}$Rb atoms strongly coupled to a weakly driven two-mode optical cavity. In the symmetry-broken phase, non-ergodic dynamics manifests in the emergence of multiple stationary states with disjoint basins of attraction. This feature enables the amplification of a small atomic population imbalance into a characteristic macroscopic cavity transmission signal. Our experiment does not only showcase strongly dissipative atom-cavity systems as platforms for probing non-trivial collective many-body phenomena, but also highlights their potential for hosting technological applications in the context of sensing, density classification, and pattern retrieval dynamics within associative memories.
\end{abstract}

\maketitle
Cavity QED systems have recently attracted considerable attention, for instance in the study of static and dynamic phase transitions \cite{Slama2007, Baumann2010, Kroeze2018, Davis2020, Schuster2020, Kessler2021, Mivehvar2021}. For atoms with multiple ground states, transitions between different levels influence the collective coupling strength \cite{Birnbaum2006, Arnold2011}, which can lead to long-lived collective dark states \cite{Orioli2022,Orioli2024,Sundar2024}, nonlinearities, and complex dynamics \cite{Suarez2023,Chu2023, Valencia2023}. Choosing hyperfine ground states, the atom-cavity coupling can even be switched on and off completely to realize controlled photon-mediated interactions between individual atoms \cite{Samutpraphoot2020}. Similarly, using clouds of atoms, so-called transmission blockade and bistability have been reported in Refs. \cite{Domokos2022, Domokos2023a}, where a cavity field was combined with a transversely irradiated laser. The same authors have theoretically shown quantum bistability in the hyperfine ground state of atoms interacting with two driven cavity modes \cite{Domokos2023b}. Moreover, multistable phases of atoms in cavities have been realized for unbalanced pumping \cite{Ferri2021}. Exploiting an intriguing link to the work by Hopfield \cite{Hopfield1982} on associative memories, atom-cavity systems have been proposed as a platform for the observation of pattern retrieval dynamics \cite{Carollo2021,marsh2021}. Here, learned patterns are encoded in the atom-light coupling constants. In the so-called pattern retrieval phase, the atomic ensemble will then dynamically evolve to the pattern that is closest to the initial state. This retrieval requires the breaking of ergodicity \cite{Fiorelli2020,Marsh2024}, i.e. the decay of configuration space into a set of disconnected parts.

\begin{figure*}[t!]
\includegraphics[width=\textwidth]{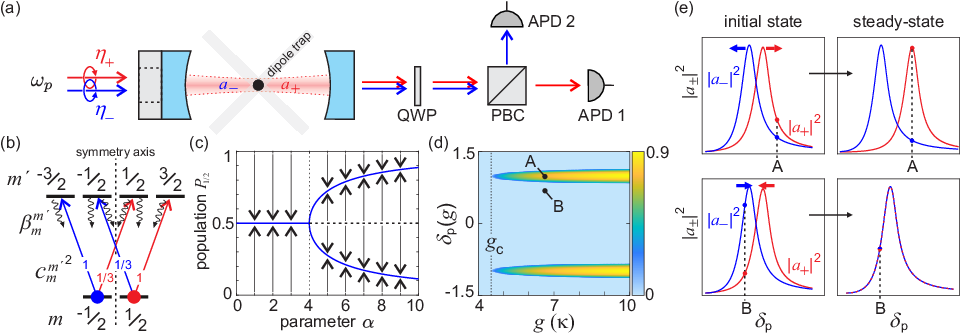}
\caption{\label{fig1}\textit{Simplified level scheme.} \textbf{(a):} Cold atoms in a cavity are pumped by two circularly polarized light beams, corresponding to cavity modes with fields $a_\pm$. The pump frequencies $\omega_p$ are detuned by $\delta_p$ from the empty cavity ans atomic resonances. The transmission of the fields through the cavity is detected on two avalanche photo diodes (APD) after separation with a quarter wave plate (QWP) and a polarizing beam cube (PBC). \textbf{(b):} Simplified level scheme consisting of two ground states and four excited states with corresponding Clebsch Gordan coefficients $c_m^{m'}$ and atomic decay branching ratios $\beta^{m'}_m$. \textbf{(c):} Analytic solutions for $P_{1/2}$ (population of the $m=1/2$ ground state) as a function of the parameter $\alpha$. A symmetry-breaking phase occurs for $\alpha\geq4$. The arrows indicate the basin of attraction to the stable solutions. \textbf{(d):} Population imbalance $I_p$ vs. the detuning $\delta_p$ and coupling strength $g$. A minimum coupling strength $g_c$ is required to observe symmetry-breaking. \textbf{(e):} Each intra-cavity power $|a_\pm|^2$ shows a symmetric normal mode splitting (only $\delta_p>0$ sketched). The initial splitting of the two modes depends of the initial atomic population imbalance (sketched $P_{1/2}>P_{-1/2}$). The choice of detuning determines whether in the steady state the population is imbalanced (upper row), or balanced (lower row).}
\end{figure*}

In this work we present evidence for a $\mathbb{Z}_2$-symmetry-breaking phase transition in an ensemble of multi-level atoms coupled to a pumped two-mode cavity (see Fig.~\ref{fig1}a). The system is strongly dissipative, i.e. the cavity modes and excited states are only weakly populated and can be adiabatically eliminated. In the symmetry-broken phase, we find evidence for non-ergodic dynamics: a population imbalance in the initial state of the atomic ensemble is converted into a macroscopic signal which can be easily detected, allowing us to monitor the phase transition by measuring the transmission of the two cavity modes. This dynamical feature can be interpreted as pattern retrieval within an associative memory \cite{Hopfield1982} or the solution of a density classification task \cite{Land1995}. The non-ergodic behaviour can, moreover, be utilized for quantum sensing applications, as recently demonstrated in Ref. \cite{Ding2022}.

\textit{Model - } We consider an ensemble of $N$ atoms with $2F+1$ ground state levels, labelled by $m=-F,-F+1,...,F$ and $2F'+1$ excited state levels, labelled by $m'=-F',-F'+1,...,F'$. The atoms interact with two circularly polarized modes of an optical cavity, represented by the creation and annihilation bosonic operators $a^\dag_l$ and $a_l$ with $l=\pm$, each of which are driven by a pump laser at a rate $\eta_l$ and frequency $\omega_p$, detuned by $\delta_p=\omega_p-\omega_a$ with respect to both cavity and atomic frequencies, $\omega_c=\omega_a$. As sketched in Fig. \ref{fig1}b, the cavity modes $a_+$ ($a_-$) drive transitions between the ground state level $m$ and the excited state $m'=m+1$ ($m'=m-1$). The Hamiltonian that describes the coherent dynamics can be decomposed into a sum $H=H_0+H_\mathrm{ia}$ of the independent cavity and atom evolution and the interaction between the two systems. In a frame rotating with the laser frequency, the independent Hamiltonian reads
\begin{equation}
\begin{split}
H_0=&\,\hbar\sum_{l=\pm}\left[-\delta_{p}a^\dagger_l a_l+\mathrm{i}\eta_l\left(a^\dagger_l-a_l\right)\right]\\
&-\hbar\delta_{p}\sum_{j=1}^N\sum_{ml}\sigma_{jm}^{m+l\dagger}\sigma_{jm}^{m+l},
\end{split}\label{eq_H0}
\end{equation}
where we have introduced the atomic ladder operator $\sigma^{m'}_{jm}=\left|m\right.\rangle_j\langle\left.m'\right|$. After the rotating wave approximation, the interaction of the atoms with the two light fields is described by
\begin{equation}\label{eq_Hia}
H_\mathrm{ia}=\mathrm{i}\hbar\sum_{j=1}^N\sum_{ml}g_{m}^{m+l}\left(a_l^\dagger\sigma^{m+l}_{jm}-a_l\sigma^{m+l \dagger}_{jm}\right).
\end{equation}
The interaction strength $g_{m}^{m'}=g_0c_{m}^{m'}$, with atom-cavity coupling constant $g_0$, depends on the transition via its corresponding Clebsch-Gordan coefficient $c_{m}^{m\pm1}$.
Including the atomic and cavity field decay, the dynamics of the system is ultimately described by the quantum master equation
\begin{equation}
\begin{split}
\dot\rho=&-\frac{\mathrm{i}}{\hbar}[H,\rho]+2\kappa\sum_{l}\left(a_l\rho a_l^{\dagger}-\frac{1}{2}\left\{a_l^{\dagger} a_l,\rho\right\}\right)\\
&+\Gamma\sum_{j=1}^N\!\sum_{mm'}\!\beta^{m'}_{m}\!\left(\!\sigma^{m'}_{jm}\rho \sigma^{m'\dagger}_{jm}-\frac{1}{2}\left\{\sigma^{m'\dagger}_{jm}\sigma^{m'}_{jm},\rho\right\}\!\right),
\end{split}\label{eq:masterequation}
\end{equation}
where $\Gamma\beta^{m'}_{m}$ is the decay rate for each atom from level $\left|m'\right>$ to $\left|m\right>$ and $2\kappa$ the one of the cavity field. The master equation exhibits a $\mathbb{Z}_2$-symmetry when $\eta_+=\eta_-$, since the values of $\beta^{m'}_{m}$ and $g_{m}^{m'}$ are symmetric under the transformation $m\to-m$, $m'\to-m'$ and $l\to -l$ (see Fig.~\ref{fig1}b).

\textit{Emergent collective dynamics -} While Eq. \eqref{eq:masterequation} is valid for an arbitrary number of atomic levels, 
the essential features of the dynamics can be understood in the simple level scheme shown in Fig.~\ref{fig1}b with $F=1/2$ and $F'=3/2$. Here, following the method outlined in Ref. \cite{Suarez2023} for a single mode cavity, we derive a rate equation for the two ground state populations $P_{\pm1/2}$ in the weak pump limit. In this limit, the population of the excited states is negligible, a condition that is satisfied by the parameters of our experiment. For equal pumping, $\eta^2=\eta_-^2=\eta_+^2$, the rate equation is given by 
\begin{equation}\label{eq:P_dot}
\dot{P}_{1/2}=-\Gamma_\mathrm{eff}\left[f(P_{1/2})P_{1/2} - f(P_{-1/2})(P_{-1/2})\right]
\end{equation}
where $P_{1/2}+P_{-1/2}=1$, and where the exact form of the effective decay rate $\Gamma_\mathrm{eff}\propto \eta^2$ and nonlinear function $f$ are given in the Supplementary Material~\cite{Supplementary}. Setting $\dot{P}_{1/2}=0$ to find the steady state value of $P_{1/2}$ leads to the cubic equation
\begin{equation}\label{eq:cubic}
2\alpha P_{1/2}^3-3\alpha P_{1/2}^2 + (2+\alpha)P_{1/2}-1=0,
\end{equation}
with a parameter $\alpha$ that depends on the pump detuning $\delta_p$ and on the (collective) coupling strength $g\equiv g_0\sqrt{N}$. Both parameters can be tuned experimentally \cite{Supplementary}. In order to allow for a generalization to larger degeneracy of the ground state manifold we introduce the population imbalance $I_p=\sum_{m>0}P_m-\sum_{m<0}P_m$ as order parameter of the phase transition, which in case with two ground state levels is given by $I_p=P_{1/2}-P_{-1/2}$. The stable steady state solution of this parameter is given by 
\begin{equation}\label{eq:P1_solution}
I_p=\begin{cases}
0 &\text{ for } \alpha\leq 4,\\
\pm\sqrt{\frac{\alpha-4}{\alpha}} &\text{ for } \alpha>4,
\end{cases}
\end{equation}
i.e. a phase transition occurs at the critical point with $\alpha_c=4$ from a symmetric phase with $P_{1/2}=P_{-1/2}=1/2$ to a symmetry-broken phase, as one can observe in Fig.~\ref{fig1}c. A third unstable steady-state solution exists for $\alpha>4$ with $\mathcal{R}e\left[P_{1/2}\right]=1/2$ and $\mathcal{I}m\left[P_{1/2}\right]<0$. Fig.~\ref{fig1}d shows that the symmetry-broken phase with $I_p\neq 0$ in the stationary state emerges in a frequency range around the normal mode splitting of the strongest transition, i.e. for $\delta_p\approx \pm g$, if the coupling strength exceeds a critical value of $g_{c}\approx 4.51 \kappa$ \cite{Supplementary}. 

This phase transition can be qualitatively understood in a simple picture sketched in Fig. \ref{fig1}e: Any imbalance in the populations of levels $m=-1/2$ and $m=1/2$ leads to a different normal mode splitting for each cavity mode. Thus, for a given fixed detuning $\delta_p$, the intracavity photon numbers $|a_\pm|^2$ are in general different, despite equal pumping. Let us now for concreteness assume that initially there is a small excess of atoms in level $m=1/2$, corresponding to the splitting shown in the first column of Fig.~\ref{fig1}e. By choosing a detuning where $|a_+|^2>|a_-|^2$ (first row in Fig.~\ref{fig1}e), more atoms will then be pumped to $m=1/2$. Thus, the normal mode splitting of level $m=1/2$, i.e. $|a_+|^2$, increases, and that of level $m=-1/2$, $|a_-|^2$, decreases. This further enhances the pumping to level $m=1/2$, which becomes a runaway process and increases in turn the population imbalance. This dynamics thus features (i) symmetry-breaking: the initial condition selects which stationary state is chosen, and (ii) leads to density classification: small population imbalances become macroscopically amplified. If, on the other hand, the detuning is set to a value where $|a_-|^2>|a_+|^2$ (second row in Fig.~\ref{fig1}e), more atoms are pumped to $m=-1/2$, thereby decreasing the excess of population. As the same argument holds for an excess of atoms in $m=-1/2$, the dynamics will approach a steady state with equal populations where $I_p=0$, and no ergodicity-breaking is observed.

\textit{Universal features -} In the following we investigate the universal properties of the phase transition, with the main results summarized here while details are found in the Supplementary Material \cite{Supplementary}. The population imbalance $I_p$ near the transition point at $\alpha_c=4$ can be written as $I_p\sim\pm(\alpha-\alpha_c)^{\beta}$, with critical exponent $\beta=1/2$. This is the same scaling behavior as the one observed in the mean-field Ising model with vanishing magnetic field $H$, where the magnetization scales like $M(T,H=0)=|T-T_c|^\beta$, with temperature $T$ and critical temperature $T_c$ \cite{Marcuzzi2014}. When weakly perturbing the system near the critical point, it relaxes to stationarity via a power-law behavior: indeed, for $\alpha=\alpha_c$ small imbalances return to the steady state as $I_p\sim\left(\Gamma_\mathrm{eff}t\right)^{-1/\zeta}$, with critical exponent $\zeta=2$. Finally, introducing an external field $\Delta\eta$ that breaks the symmetry of the two stable solutions by the relative imbalance in the pumping rates, $\Delta\eta=(\eta_+^2-\eta_-^2)/\eta_-^2$, we obtain that the resulting imbalance scales as 
$I_p\sim \Delta\eta^{1/\delta}$, with critical exponent $\delta=3$. We conclude that for two ground states the universality class associated with the phase transition is the same as the one of the mean-field Ising model.\\

\begin{figure}[t]
\includegraphics[width=\linewidth]{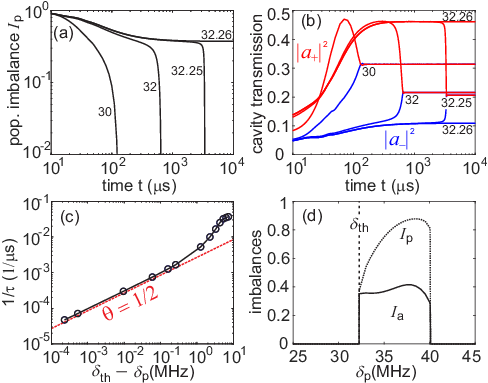}
\caption{\label{fig2}\textit{Simulation of the dynamics.} The atom number is set to $N=2\times10^{4}$ with $g_0=0.0654\kappa$, close to the values of the experiment. All simulations start at $t=0$ with $P_{2}=1$ and $P_m=0$ for $m\neq2$. Dynamics of \textbf{(a):} the population imbalance $I_p=P_{2}+P_{1}-P_{-1}-P_{-2}$ and \textbf{(b):} the intracavity power, for various values of detuning $\delta_p$ (in MHz). In the symmetry-broken regime, for $\delta_p>\delta_\mathrm{th}\approx 32.259$~MHz, at very long times the population imbalance remains $I_p\neq 0$ and the cavity fields do not reach the same value. \textbf{(c):} Close to the transition point $\delta_\mathrm{th}-\delta_p\ll 1$ the relaxation time $\tau$, defined here as the time it takes for $I_p$ to decrease below $0.1$, diverges algebraically with slope $\theta=1/2$. \textbf{(d):} The symmetry-broken phase is characterized by a steady state imbalance of both the ground state populations $I_p$ and the transmitted light power $I_a=|a_+|^2-|a_-|^2$.} 
\end{figure}

\begin{figure*}[t!]
\includegraphics[width=\textwidth]{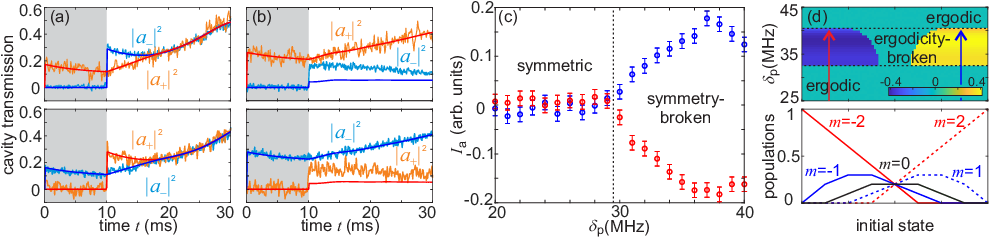}
\caption{\label{fig3}\textit{Symmetry breaking.} Cavity transmission in the symmetric phase for \textbf{(a):} $\delta_p=23$~MHz, and within the symmetry-broken phase for \textbf{(b):} $\delta_p=33$~MHz, and comparison with  simulations (solid lines). The populations are prepared in an unbalanced state by 10~ms of initial pumping (grey shaded area). Top and bottom subfigures correspond to an inverted time sequence of the pump beams. In (a) the final state does not depend on the initial conditions whereas in (b) the ergodicity is broken and the system evolves to one of the two solutions, depending on the time sequence. \textbf{(c):} Cavity field imbalance $I_a$ measured at $t=20~$ms as a function of the detuning, where the transition from the symmetric to the symmetry-broken phase is clearly visible at $\delta_p\approx30$~MHz. Blue and red data correspond to an inverted pump time sequence. \textbf{(d):} The upper panel shows the numerically simulated $I_a$ as a function of the detuning $\delta_p$ and the initial populations $P_m$, shown in the lower panel. Unlike for the simple model with only two ground states, a stationary state with $I_a=0$ (green area) exists for the same detuning as those with $I_a\neq 0$ (yellow and blue). The red and blue arrow indicate the directions along which we experimentally probe the phase transition in (c).}
\end{figure*}

\textit{Numerical simulation -} We consider now a more complex case, where each atoms has five and seven ground and excited state levels, respectively ($F=2$ and $F'=3$), which reflects the situation in our experiment. The level scheme, including the values of $c_m^{m'}$ and $\beta_m^{m'}$, is given in \cite{Supplementary}. Here, the large amount of levels makes deriving and analysing closed equations such as Eq. \eqref{eq:P_dot} infeasible. We use instead a fully numerical approach: we derive and solve the mean-field Heisenberg equations of motion starting from the master equation \eqref{eq:masterequation} with help of the QuantumCumulants package of the Julia software \cite{Plankensteiner2022}. The pumping strength is chosen as $\eta_+=\eta_-=14\kappa$, which is large enough to keep the simulations fast, and small enough to avoid saturation of the atoms (excited state population stays well below $1\%$ for all times). The initial populations are set to $P_{2}=1$ and $P_m=0$ for $m\neq2$ to put the system in the maximally imbalanced state. The Heisenberg equations are numerically integrated to obtain the time evolution until the steady state is reached. Results are shown in Fig.~\ref{fig2}a and b for the population imbalance $I_p=P_{2}+P_{1}-P_{-1}-P_{-2}$ and cavity transmission $|a_\pm|^2$, respectively, for different values of the detuning $\delta_p$. We observe that for all values $\delta_p< \delta_\mathrm{th}\approx 32.259$ MHz the imbalance both in atomic population and cavity transmission eventually reach a value of zero, while for $\delta_p>\delta_\mathrm{th}$ the imbalance persists in both quantities for all times. Close to the phase transition, a metastable plateau is observed \cite{Marcuzzi2014}. Fig.~\ref{fig2}c shows that the time $\tau$ it takes until the imbalance vanishes diverges algebraically at the phase transition, i.e. $1/\tau\propto(\delta_\mathrm{th}-\delta_p)^\theta$ with exponent $\theta=1/2$. These results are summarized in Fig.~\ref{fig2}d, where we clearly identify the symmetry-broken region as the detuning changes.

\textit{Experimental results -} To probe the phase transition and ergodicity-breaking behaviour we use ultracold $^{87}$Rb atoms, which are cooled in a magneto-optic trap and loaded into a crossed beam optical dipole trap that is overlapped with the mode volume of an optical cavity, as depicted in Fig.~\ref{fig1}a. Further details on the experimental preparation are described in \cite{Supplementary}. We initially prepare ground state populations with a strong imbalance by $10$~ms of optical pumping with only one of the two cavity fields. After the pumping, the second cavity field is switched on with equal pump strength, i.e. $\eta_+=\eta_-$, and we observe the dynamics of the transmitted light powers on the corresponding APDs. A series of measurements for detunings $\delta_p$ between 20~MHz and 40~MHz is shown in the Supplementary Material \cite{Supplementary}. We focus here on the first $30$~ms, where the signatures of the different phases become apparent. 
The scaling and deviations between the experimental data and simulations including atom loss are explained in \cite{Supplementary}. Here, we detail two typical measurements within and outside the symmetry-broken regime. Figs.~\ref{fig3}a and b show the dynamics within the symmetric phase and the symmetry-broken regime, respectively. The smoking gun for symmetry-breaking is here the observation that the transmission curves do not tend to the same value. Top and bottom subfigures show the dynamics when the time sequence of the two pumps is inverted, such that 
the initial population at $t=10$~ms is imbalanced in favor of states with $m<2$ or $m>2$, respectively. This has no effect on the steady state in the symmetric phase. In contrast, in the symmetry-broken phase the subsequent dynamics is exchanged, signalling the breaking of ergodicity.

The phase transition also manifests in the imbalance $I_a=|a_+|^2-|a_-^2|$ of the cavity fields at long times as shown by the experimental data displayed in Fig.~\ref{fig3}c as a function of the detuning. We have chosen a time of $t=20$~ms for this analysis, where 
the effect of atom loss is still negligible. We observe a non-zero imbalance for detunings $\delta_p\gtrsim 30$~MHz, whose sign depends on the order of the pump time sequence. Determining the value of the threshold more precisely is prevented by the divergence of time scales close to the phase transition and the subsequent influence of atom losses: Here, starting in the symmetric phase closely below the threshold, atom losses can cause the transition into the symmetry-broken phase and vice versa as the reduction of $N$ leads to a decreased $g=g_0\sqrt{N}$, as shown in \cite{Supplementary}. Fig.~\ref{fig3}d shows
the simulated field imbalance $I_a$ in the steady state as function of the detuning for various initial populations. To simplify the analysis, we project the four-dimensional parameter space of the ground-state levels to a line (plotted on the lower axis) that connects the fully stretched states and contains the state with equal populations at its center. We observe again that, for a certain range of detunings, ergodicity is broken. In the previously considered case of atoms with two ground states, three stationary states exist out of which, however, only two are stable (Fig.~\ref{fig1}c). Here, conversely, three different stable basins of attraction exist, and the initial conditions determine which one of the stationary states is reached. This is reminiscent of the multi-critical Ising model, which is discussed in Ref. \cite{Overbeck2017}, and suggests that the transition between symmetric and symmetry-broken phase may be governed by a different universality class. In our experiment, however, we could not probe the central region in which $I_a=0$ as this required a fine-tuning of the initial populations beyond our experimental control.

\textit{Conclusion -}  We have observed a symmetry-breaking phase transition in a strongly dissipative system composed of a driven two-mode cavity and an ensemble of multi-level atoms. In the symmetry-broken phase, several stationary states exist each with their own basin of attraction. This setting bears close resemblance with density classification dynamics, or pattern retrieval within a Hopfield associative memory. In this context, it would be interesting to move away from the strongly dissipative limit. This is expected to lead to the emergence of new (quantum) patterns which have been predicted in quantum generalized Hopfield neural networks \cite{Rotondo2018}. Moreover, multi-state atoms allow to engineer and probe even more exotic models, such as quantum generalized Potts-Hopfield neural networks \cite{Fiorelli2022}. We also plan to augment our experiment by including competing long-range interactions between neighboring atoms using Rydberg dressing \cite{Glaetzle2015}. Another future perspective is to generate and utilize atom number squeezing in the symmetric phase \cite{Braverman2019}.

\begin{acknowledgments}
The project was funded by the Deutsche Forschungsgemeinschaft (DFG, German Research Foundation) - 422447846, 465199066 and 465199066. It was carried out within research unit FOR 5413 "Long-range interacting quantum spin systems out of equilibrium: Experiment, Theory and Mathematics". IL further acknowledges funding through the Deutsche Forschungsgemeinschaft (DFG, German Research Foundation), through project 449905436 and the Machine Learning Cluster of Excellence under Germany’s Excellence Strategy --- EXC number 2064/1 – Project number 390727645.
\end{acknowledgments}

\bibliography{references}

\newpage
\section{Supplementary Material}
\subsection{Derivation of rate equation \eqref{eq:P_dot}}
We derived in \cite{Suarez2023} a rate equation for $P_{-1/2}$ where a single mode drives the transition from $m=-1/2$ to $m=1/2$. A corresponding rate equation exists for $P_{1/2}$ where the second mode drives the transition from $m=1/2$ to $m=-1/2$:
\begin{eqnarray}
\dot{P}_{-1/2}&=&-\Gamma_\mathrm{eff}^+f^+(P_{-1/2})P_{-1/2},\\
\dot{P}_{1/2}&=&-\Gamma_\mathrm{eff}^-f^-(P_{1/2})P_{1/2}.
\end{eqnarray}
As $P_{-1/2}+P_{1/2}=1$, both equations can be summarized as 

\begin{equation}\label{eq:P_dot_suppl}
\dot{P}_{-1/2}=-\Gamma_\mathrm{eff}^+f^+(P_{-1/2})P_{-1/2} +\Gamma_\mathrm{eff}^-f^-(P_{1/2})P_{1/2},
\end{equation}
where the first term describes loss of $P_{-1/2}$ due to pumping to level $m={1/2}$ by field $a_+$ and the second term describes gain of $P_{-1/2}$ due to pumping from level $m={1/2}$ by field $a_-$, with corresponding effective pumping rates
\begin{eqnarray}\label{eq:Gamma_eff_p_suppl}
\Gamma_\mathrm{eff}^+=&\frac{c_{1+}^2\eta_+^2}{g_0^2N^2}\frac{\beta^2_2\Gamma}{c_{2+}^2\left(c_{2+}^2+u\right)+w},\\\label{eq:Gamma_eff_m_suppl}
\Gamma_\mathrm{eff}^-=&\frac{c_{2-}^2\eta_-^2}{g_0^2N^2}\frac{\beta^1_1\Gamma}{c_{1-}^2\left(c_{1-}^2+u\right)+w}.
\end{eqnarray}
Non-linearity is generated by the function
\begin{equation}
\label{eq:f(P)_suppl}
f^\pm(P_m)=\frac{1}{\alpha_\pm P_{m}^2+\beta_\pm P_{m} + 1},
\end{equation}
with parameters
\begin{eqnarray}
\label{eq:alpha_p_suppl}
\alpha_+=&\frac{\left(c_{1+}^2-c_{2+}^2\right)^2}{c_{2+}^2\left(c_{2+}^2+u\right)+w},\\ \label{eq:alpha_m_suppl}
\alpha_-=&\frac{\left(c_{2-}^2-c_{1-}^2\right)^2}{c_{1-}^2\left(c_{1-}^2+u\right)+w},
\end{eqnarray}
and 
\begin{eqnarray}
\label{eq:beta_p_suppl}
\beta_+=&\frac{2\left(c_{1+}^2-c_{2+}^2\right)\left[c_{2+}^2+\frac{u}{2}\right]}{c_{2+}^2\left(c_{2+}^2+u\right)+w},\\ \label{eq:beta_m_suppl}
\beta_-=&\frac{2\left(c_{2-}^2-c_{1-}^2\right)\left[c_{1-}^2+\frac{u}{2}\right]}{c_{1-}^2\left(c_{1-}^2+u\right)+w},
\end{eqnarray}
where
\begin{align}
\label{eq:u_suppl}
u=&\frac{\Gamma\kappa-2\Delta_a\Delta_c}{g_0^2N},\\ \label{eq:w_suppl}
w=&\frac{\left[\left(\frac{\Gamma}{2}\right)^2+\Delta_a^2\right]\left[\kappa^2+\Delta_c^2\right]}{g_0^4N^2}.
\end{align}
Note that in the present work the atom detuning $\Delta_a=\omega_p-\omega_a$ equals the cavity detuning $\Delta_c=\omega_p-\omega_c$ for which $\Delta_a\Delta_c=\Delta_a^2=\Delta_c^2=\delta_p^2$.  
Due to the symmetry of the Clebsch Gordan coefficients with $c_{1-}^2=c_{2+}^2$, and $c_{1+}^2=c_{2-}^2$ and of the branching ratios with $\beta^1_1=\beta^2_2$, the effective pumping rates are $\Gamma_\mathrm{eff}=\Gamma_\mathrm{eff}^+=\Gamma_\mathrm{eff}^-$ with
\begin{equation}\label{eq:Gamma_eff2_suppl}
\Gamma_\mathrm{eff}=\frac{c_{1+}^2\eta_\pm^2}{g_0^2N^2}\frac{\beta^2_2\Gamma}{c_{2+}^2\left(c_{2+}^2+u\right)+w}.
\end{equation}
Also, because of symmetry, $\alpha\equiv\alpha_+=\alpha_-$, and $\beta\equiv\beta_+=\beta_-$, and the function $f(P_m)\equiv f^+(P_m)=f^-(P_m)$:
\begin{eqnarray}
\label{eq:alpha_suppl}
\alpha=&\frac{\left(c_{1+}^2-c_{2+}^2\right)^2}{c_{2+}^2\left(c_{2+}^2+u\right)+w},\\
\label{eq:beta_suppl}
\beta=&\frac{2\left(c_{1+}^2-c_{2+}^2\right)\left[c_{2+}^2+\frac{u}{2}\right]}{c_{2+}^2\left(c_{2+}^2+u\right)+w},\\
\label{eq:f_suppl}
f(P_m)=&\frac{1}{\alpha P_{m}^2+\beta P_{m} + 1}.
\end{eqnarray}
Thus, Eq. \eqref{eq:P_dot_suppl} can be written as
\begin{equation}\label{eq:P_dot_final_suppl}
\dot{P}_{-1/2}=-\Gamma_\mathrm{eff}\left[f(P_{-1/2})P_{-1/2} - f(P_{1/2})P_{1/2}\right],
\end{equation}
and
\begin{equation}\label{eq:P_dot_final2_suppl}
\dot{P}_{1/2}=-\Gamma_\mathrm{eff}\left[f(P_{1/2})P_{1/2} - f(P_{-1/2})P_{-1/2}\right].
\end{equation}

\subsection{Relaxation dynamics at the phase transition}
We start with the differential equation \eqref{eq:P_dot} at the phase transition with $\alpha=4$, equal pumping $\eta_+^2=\eta_-^2$, and replace $P_{1/2}$ by $P_{1/2}+\Delta P$. This results in a differential equation for the deviation
\begin{equation}\label{eq:deltaP_dot_suppl}
\Delta \dot{P}=-\Gamma_\mathrm{eff}\frac{32\Delta P^3}{64\Delta P^4-(4\beta^2+16\beta)\Delta P^2+(\beta+4)^2},
\end{equation}
that can be integrated to 
\begin{equation}\label{eq:t(P)}
\Gamma_\mathrm{eff}t=-\frac{\Delta P^2}{2}+\frac{\beta(\beta+4)}{8}\ln(\Delta P)+\frac{(\beta+4)^2}{16\Delta P^2}+C,
\end{equation}
with integration constant $C$ determined by the initial population. For small deviations $\Delta P\ll1$ from the steady state and for $\beta\neq-4$ , the $1/\Delta P^2$ term dominates, and the solution $\Delta P(t)$ is a power law
\begin{equation}\label{eq:P(t)}
\Delta P(t)=\frac{4}{\beta+4}\left(\Gamma_\mathrm{eff}t\right)^{-1/2}.
\end{equation}
As the population deviation is connected with the population imbalance as $I_p=2\Delta P$, the dynamics of $I_p(t)$ follows the same power law.

\subsection{Unbalanced pumping at the phase transition}
We start with differential equation \eqref{eq:P_dot} and express unbalanced pumping by introducing 
$d\eta^2=\eta_+^2-\eta_-^2$. Thus,  
$\Gamma_\mathrm{eff}^+=\Gamma_{e}\eta_-^2+\Gamma_{e}d\eta^2$, with 
$\Gamma_e=\Gamma_\mathrm{eff}^\pm/\eta_\pm^2$, and we get 
\begin{eqnarray}\label{eq:P_dot_unbal_supp}
\dot{P}_{1/2}=&-&\Gamma_e\eta_-^2\left(f(P_{1/2})P_{1/2} - f(1-P_{1/2})(1-P_{1/2})\right)\nonumber\\
          &-&\Gamma_{e}d\eta^2f(P_{1/2})P_{1/2}.
\end{eqnarray}
We insert the function $f$ from \eqref{eq:f_suppl}, set $\alpha=4$ to operate at the critical point and calculate the steady state by setting  $\dot{P}_{1/2}=0$ in \eqref{eq:P_dot_unbal_supp}. We again introduce the population deviation $\Delta P=P_{1/2}-1/2$ from the balanced steady state solution at the critical point. This leads to an algebraic equation for $\Delta P$:
\begin{eqnarray}\label{eq:unbal01_suppl}
-&\eta_-^2&\frac{32\Delta P^3}{64\Delta P^4-(4\beta^2+16\beta)\Delta P^2+(\beta+4)^2}-\nonumber\\
-&d\eta^2&\frac{2\Delta P+1}{8\Delta P^2+2(\beta+4)\Delta P+(\beta+4)}=0,
\end{eqnarray}
which further leads to a 5th order polynomial equation:
\begin{eqnarray}\label{eq:unbal02_suppl}
&128&\left(2+\Delta\eta\right)\Delta P^5 + 64\left(\beta+4+\Delta\eta\right)\Delta P^4+\nonumber\\
+&8&(\beta+4)\left(4-\beta \Delta\eta\right)\Delta P^3 - 4\Delta\eta\beta(\beta+4)\Delta P^2 + \nonumber\\
+&2&\Delta\eta(\beta+4)^2\Delta P +\Delta\eta(\beta+4)^2=0
\end{eqnarray}
where we introduced the relative unbalanced pumping $\Delta\eta=\frac{d\eta^2}{\eta_-^2}$. We solve this equation up to third order in $\Delta P\ll1$, which is the lowest order that delivers an $\Delta\eta$-dependent solution $P(\Delta\eta)$, thus we get the equation
\begin{equation}\label{eq:unbal03_suppl}
8\left(4-\beta \Delta\eta\right)\Delta P^3 - 4\beta \Delta\eta\Delta P^2 +2(\beta+4)\Delta\eta \Delta P +(\beta+4)\Delta\eta=0,
\end{equation}
which we solve for $\Delta\eta$:
\begin{eqnarray}\label{eq:unbal04_suppl}
\Delta\eta&=&\frac{-32\Delta P^3}{-8\beta \Delta P^3-4\beta \Delta P^2+2(\beta+4)\Delta P+(\beta+4)}\nonumber\\
&\approx&-\frac{32\Delta P^3}{\beta+4},
\end{eqnarray}
where in the second step the denominator was approximated for small deviations $\Delta P$.
Thus, we can solve for the population deviation and get
\begin{equation}\label{eq:unbal05_suppl}
\Delta P(\Delta\eta)=-\frac{1}{2}\left(\frac{\beta+4}{4}\right)^{1/3}\Delta\eta^{1/3}.
\end{equation}
Thus, the population difference at the critical point scales like $\Delta\eta^{1/\delta}$ with critical exponent $\delta=3$. 

\subsection{Critical coupling strength}
The critical value $\left(g_0\sqrt{N}\right)_c$ of the collective coupling strength is derived from  Eq.~\eqref{eq:alpha_suppl} by setting $\alpha=4$ and $\delta_p=g_0\sqrt{N}$:
\begin{equation}\label{eq:g_cr1_suppl}
\frac{\left(c_{1+}^2-c_{2+}^2\right)^2}{c_{2+}^2(c_{2+}^2+\frac{\Gamma\kappa-2g_0^2N}{g_0^2N})+\frac{(\Gamma^2/4+g_0^2N)(\kappa^2+g_0^2N)}{g_0^4N^2}}=4
\end{equation}
This is a quadratic equation
\begin{equation}
    ax^2+bx+c=0
\end{equation}
in $x=g_0^2N$ with the coefficients
\begin{eqnarray}
a&=&\left[\frac{1}{4}\left(c_{1+}^2-c_{2+}^2\right)^2-\left(c_{2+}^2-1\right)^2\right]\nonumber\\
b&=&\left[c_{2+}^2\Gamma\kappa+\frac{\Gamma^2}{4}+\kappa^2\right]\nonumber\\
c&=&-\frac{\Gamma^2}{4}\kappa^2.
\end{eqnarray}
For the present case with two ground state levels with $c_{1+}^2=1/3$, $c_{2+}^2=1$, and $\Gamma=\kappa$, we get 
\begin{equation}
    \frac{x}{\kappa^2}=\frac{9}{8}\left(9+\frac{\sqrt{745}}{3}\right)\approx20.36,
\end{equation}
corresponding to $\left(g_0\sqrt{N}\right)_c\approx4.51\kappa$.

\subsection{Experimental details}
\begin{figure}[t]
\includegraphics{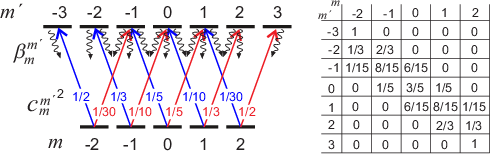}
\caption{\label{fig1_suppl}\textbf{Experimental level scheme:} The five Zeeman levels of the $5S_{1/2},F=2$ ground state of $^{87}$Rb are coupled to the seven Zeeman levels of the $5P_{3/2}$ excited state. The table indicates the branching ratios $\beta_m^{m'}$ of the decay from excited level $m'$ to ground state level $m$. Transitions that are driven by field $a_+$ are plotted in red, those driven by field $a_-$ are plotted in blue.} 
\end{figure}
We work with $^{87}$Rb atoms, see Fig.~\ref{fig1_suppl}. The atoms are collected in a magneto-optic trap and prepared in the $5S_{1/2},F=1$ ground state after Sisyphus cooling to $T=20~\mu K$, while being loaded into a crossed beam optical dipole trap (wavelength $\lambda=1064~$nm and power $P=6~$W). We measure an AC Stark shift in the dipole trap of $U_\mathrm{AC}=1.9$~MHz by spectroscopy of the trapped atoms using a free space probe beam transverse to the cavity. The dipole trap is overlapped with the mode volume of an optical standing wave cavity with cavity length $l=5~$cm and Finesse $F=220$. The length of the cavity is stabilized by the Pound-Drever-Hall (PDH) technique to a laser with wavelength $\lambda=786~$nm (the lock laser), sufficiently far detuned from the D2 line to avoid noticeable interaction with the atoms. Thus, the cavity is tuned in resonance with the atomic transition, including the AC Stark shift caused by the dipole trap. The lock laser itself is stabilized by another PDH lock to an ultrastable ULE cavity. The two cavity pump lasers (pump 1 and 2) are each stabilized by a beat lock to a reference laser which is locked to Rubidium by saturated absorption spectroscopy. The two pump lasers are individually switched by AOMs, overlapped with orthogonal polarization on a polarizing beam splitter and sent together through a polarization-maintaining optical single mode fiber to guarantee optimum mode overlap. After the fiber, some portion of the light is outcoupled using a non-polarizing beam splitter to check the relative powers of the two pump fields. Then, a quarter wave plate generates the corresponding circular polarization. Then the beams are coupled into the cavity. At the transmission, we distinguish both beams using another quarter wave plate and a polarizing beam splitter, detecting them separately on two avalanche photodiodes after filtering the lock laser from the signals using an interference filter. Prior to the experiment, the number of atoms in the dipole trap is tuned by extending the time they remain trapped. After several seconds, the initial number of atoms coupled to the cavity pump field is $N\sim 20\,000$. A re-pump field is activated $100$~ms before the cavity pump laser, pumping the atoms back to the $5S_{1/2},F=2$ ground state, and kept on during the rest of the experiment.
\begin{figure}[t]
\includegraphics[width=\linewidth]{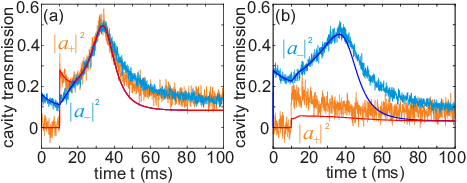}
\caption{\label{fig2_suppl}\textbf{Dynamics for an imbalanced initial state:} Measured cavity transmission (a) outside the bistability region for $\delta_p=23$~MHz and (b) within the bistability region for $\delta_p=33$~MHz. The simulations (solid lines) including atom loss reproduce the experimental curves.} 
\end{figure}
\subsection{Dynamics including atom loss}
\begin{figure}[t]
\includegraphics[width=0.7\linewidth]{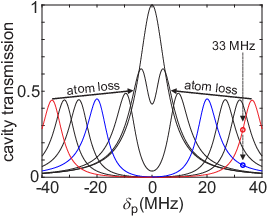}
\caption{\label{fig3_suppl}\textbf{Normalization of cavity transmission:} Atom loss reduces the normal mode splitting. Thus, the transmission crosses a maximum with defined height if the normal mode is probed at the inner slope as in the red curve. This allows us to re-gauge the transmission.} 
\end{figure}

In the main part of the paper we have shown plots for the first 30~ms of the cavity transmission dynamics, where atom loss is less than 10\%. There, the effects of the different phases become apparent. In Fig.~\ref{fig2_suppl} we show the full time dynamics up to 100~ms. In a) the internal state dynamics is limited to the first $20$~ms, until both transmissions reach the same value, i.e. the steady state. The subsequent dynamics, i.e. the peak observed at $t=45$~ms and the decay after the peak, is caused by atom loss from the dipole trap due to light scattering from the intracavity light, see also Fig.~\ref{fig3_suppl}. The peak is used for normalizing the transmission, see section on normalizing the cavity transmission. The experimental data agree quantitatively with a simulation where we heuristically include atom loss by an analytical formula $N(t)=1-\left(1+e^{-\frac{t-t0}{\Delta t}}\right)^{-1}$.  In order to find good agreement between simulation and experimental data, we fit the initial ground state populations $P_m$, the pump rate $\eta_+=\eta_-$ and the time $t_0$ and duration $\Delta t$ of the atom decay. The fact that the simulated curves tend to zero more quickly for large times than the experimental data, is caused by an overestimation of atom loss by the analytical $N(t)$.

\subsection{Normalizing the cavity transmission}
\label{sec:normalization}
We initially scale the height of the measured cavity transmission to the maximum of the measured empty cavity resonance, corresponding to the procedure in the simulations. We observe that the transmission with atoms loaded in the cavity is less than expected for the mode with larger intra-cavity power and more than expected for the mode with less intra-cavity power. We attribute this observation to a changed in-coupling efficiency into the cavity for the two modes. This is caused firstly by light scattering from the atoms into free space. This additional cavity loss reduces the incoupling efficiency of both modes equally. Secondly, birefringence of the cavity can influence the incoupling efficiency differently. Although no birefringence is observed in transmission if only a single mode is pumped, interference at the in-coupling mirror can enhance this effect if both modes are pumped. Such in-coupling effects are effectively changing the pump rates $\eta_\pm$, and are not taken into account in the simulations. We thus have to re-normalize the curves. For modes that are probed at the inner slope of the normal mode splitting atom loss reduces the normal mode splitting which leads to the observation of a maximum with characteristic height for each detuning,  Fig.~\ref{fig3_suppl}. Note that the same is true if the coupling strength reduces because of pumping between the Zeeman sublevels. We scale the maximum of the measured transmission curve to this value. Thus, in Fig. \ref{fig2_suppl}a) of the Supplementary [Fig.~3a) in the paper] the experimental curves were both scaled by a factor of $1.46$. Modes that are probed at the outer slope do not feature such a maximum. This is the case for the orange data shown in Fig.~\ref{fig2_suppl}b) of the Supplementary. We decided that we scale those data with the same factor as the simultaneously recorded other mode (blue data in the same figure), even if we then overestimate the actual transmission. Thus, in Fig. \ref{fig2_suppl}b) of the Supplementary [Fig. 3b) in the paper] the experimental curves were both scaled by a factor of $1.62$.

\begin{figure}[t]
\includegraphics[width=0.7\linewidth]{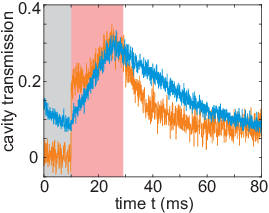}
\caption{\label{fig4_suppl}\textbf{Quench dynamics} for $\delta_p=28$~MHz: After prepumping (grey shaded region) the dynamics starts in the symmetric phase (pink shaded region) and equlibrates the transmission. Then, at the end of the pink region, the transmissions start to diverge from each other, corresponding to the symmetry-broken phase.} 
\end{figure}
\subsection{Including atom loss in the simulation}
\label{sec:atom_loss}
Atoms in the experiment are trapped in a dipole trap overlapped with the cavity mode. The temperature of the trapped atom cloud increases due to near-resonant light scattering from the cavity fields. As long as the temperature of the cloud is much smaller than the trap depth, all atoms remain trapped. However, when the temperature grows to a value comparable with the trap depth, some fraction of the atoms acquires an energy, given by the Boltzmann factor, higher than the trap depth. These atoms can leave the trap, and thus the number $N(t)$ of atoms interacting with the cavity decreases. As the number of atoms influences the intra-cavity light power which, in turn, determines the heating rate, simulating the atom number dynamics would require to solve a differential equation for $N(t)$, together with the master equation \eqref{eq:masterequation}. We want to avoid this complication and describe the atom number heuristically with a sigmoid analytical function $N(t)=1-\left(1+e^{-\frac{t-t0}{\Delta t}}\right)^{-1}$. Here, $t_0$ describes the time when half of the atoms are lost, and $\Delta t$ is the width of the sigmoid decrease. This description is working well for the initial atom loss, but overestimates the loss for large times.
\begin{table*}[t!]
                \begin{tabular}{|p{2cm} | p{1cm} | p{4cm} | p{2cm} | p{2cm} | p{1cm} |}

                    \hline

                    \textbf{Fig} & \textbf{$\eta^2$} & \textbf{$P_2/P_1/P_0/P_{-1}/P_{-2}$} & \textbf{$t_0$} & \textbf{$\Delta t$}  \\ \hline
                    3a top & 5.0e-2 & 0.29/0.20/0.17/0.16/0.18 & 37.4~ms & 7.6~ms \\ \hline
                    3a bottom& 5.0e-2 & 0.18/0.16/0.17/0.20/0.29 & 35.0~ms & 8.0~ms \\ \hline
                    3b top & 6.1e-2 & 0.76/0.17/0.07/0/0 & 51.0~ms & 10.3~ms \\ \hline
                    3b bottom& 6.8e-2 & 0/0/0.05/0.15/0.80 & 51.0~ms & 10.3~ms \\

                    \hline
                \end{tabular}
  \caption{Fit parameters}
  \label{tab1}
\end{table*}

\subsection{Atom loss induced phase transition}
As explained in the main part of the paper, atom losses can trigger the phase transition if the system is initialized close to the critical values. Here, we start at a detuning of $\delta_p=28$~MHz, below the threshold to the symmetry-broken phase. Thus, at first, the transmission reach the same steady-state value. While the atom number is being reduced due to atom loss from the trap, the threshold detuning is decreasing. Thus, suddenly the system is quenched into the symmetry-broken phase, and the transmissions start to separate from each other. This is shown in Fig.~\ref{fig4_suppl} 

\subsection{Fit parameters}
The fit parameters of the simulations shown in Fig.3 are summarized in TABLE \ref{tab1}. Populations are taken at the beginning of the symmetric pumping at $t=10$~ms.

\subsection{Complete set of measurements}
While in the main part of the paper we concentrate on two typical situations within and without the symmetry-broken region, here we show all the measurements taken for detunings between $\delta_p=20$~MHz and $\delta_p=40$~MHz.
\begin{figure*}
  \includegraphics[width=\textwidth]{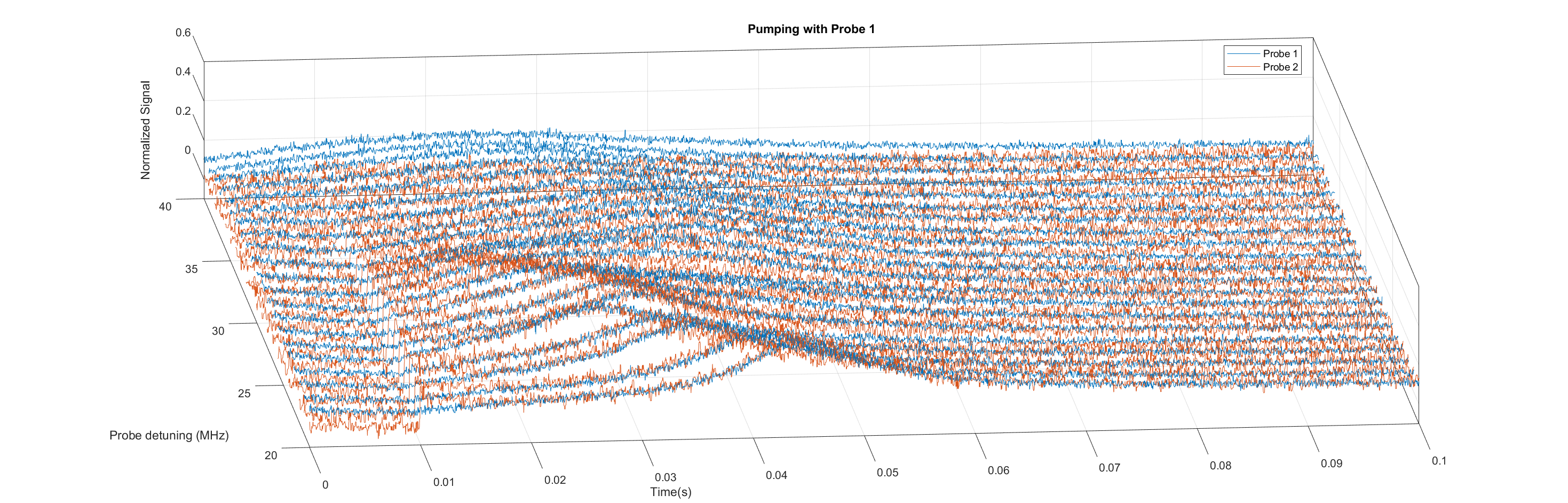}
  \includegraphics[width=\textwidth]{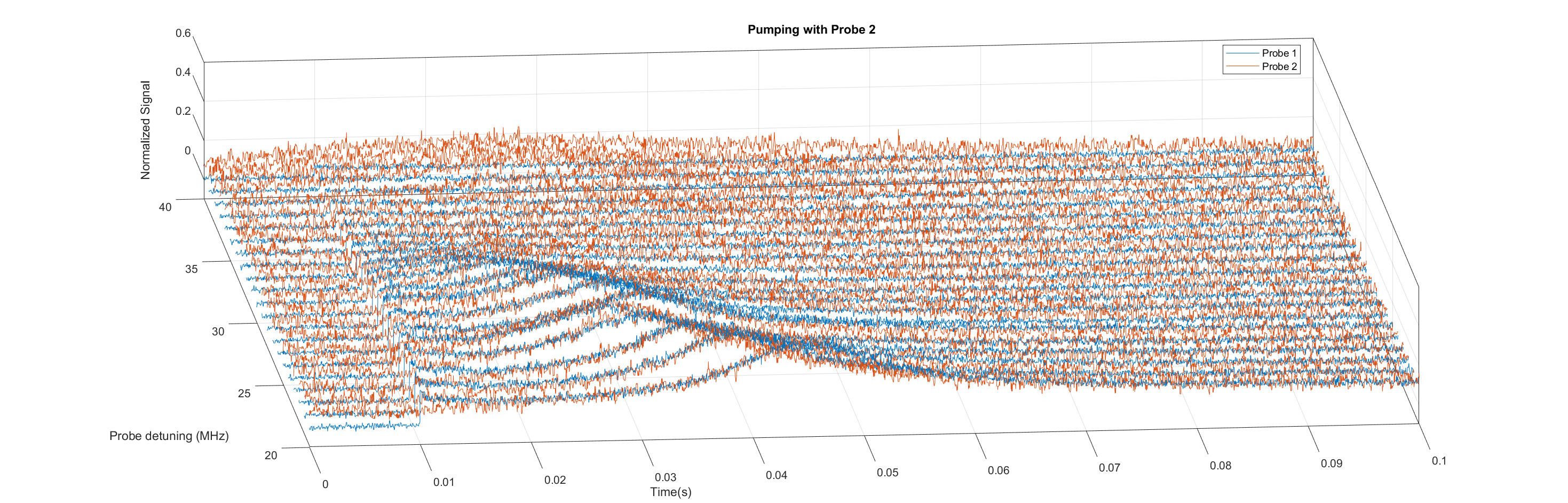}
  \caption{The data show a transition from the symmetric phase for $\delta_p\lesssim 30$~MHz to the symmetry-broken phase for $\delta_p\gtrsim 30$~MHz.}
\end{figure*}

\end{document}